\documentclass[sn-mathphys,Numbered]{sn-jnl}

\usepackage{etoolbox}   
\makeatletter
\patchcmd{\@maketitle}{\artauthors}{{\artauthors}}{}{}
\makeatother

\usepackage{graphicx}%
\usepackage{multirow}%
\usepackage{amsmath,amssymb,amsfonts}%
\usepackage{amsthm}%
\usepackage{mathrsfs}%
\usepackage[title]{appendix}%
\usepackage{xcolor}%
\usepackage{textcomp}%
\usepackage{manyfoot}%
\usepackage{booktabs}%
\usepackage{algorithm}%
\usepackage{algorithmicx}%
\usepackage{algpseudocode}%
\usepackage{listings}%
\usepackage{mathtools}

\usepackage[utf8]{inputenc}

\usepackage{stmaryrd} 

\theoremstyle{thmstyleone}%
\theoremstyle{thmstyletwo}%

\theoremstyle{thmstylethree}%

\begin{document}

\mathchardef\mhyphen="2D

\newcommand{\boldtheta}{\theta} 
\newcommand{\Ndata}{N_\mathrm{data}}
\newcommand{\Nesctheta}{N^\mathrm{esc}_\boldtheta}
\newcommand{\Vb}{V_\mathrm{b}}
\newcommand{\ntheta}{n^\mathrm{ }_\theta}
\newcommand{\phitheta}{\phi^\mathrm{ }_\theta}

\title[Article Title]{Shot-noise-induced lower temperature limit of the nonneutral plasma parallel temperature diagnostic}


\author*[1,2]{\fnm{Adrianne} \sur{Zhong}}\email{adrizhong@berkeley.edu}
\author[1]{\fnm{Joel} \sur{Fajans}}
\author[1]{\fnm{Jonathan S.} \sur{Wurtele}}

\affil[1]{\orgdiv{Department of Physics}, \orgname{University of California, Berkeley}, \orgaddress{\street{366 Physics North}, \city{Berkeley}, \postcode{94702}, \state{CA}, \country{US}}}
\affil[2]{\orgdiv{Redwood Center for Theoretical Neuroscience}, \orgname{University of California, Berkeley}, \orgaddress{\street{366 Physics North}, \city{Berkeley}, \postcode{94702}, \state{CA}, \country{US}}}


\abstract{We develop a new algorithm to estimate the temperature of a nonneutral plasma in a Penning-Malmberg trap. The algorithm analyzes data obtained by slowly lowering a voltage that confines one end of the plasma and collecting escaping charges, and is a maximum likelihood estimator based on a physically-motivated model of the escape protocol presented in \cite{beck1990thesis}. Significantly, our algorithm may be used on single-count data, allowing for improved fits with low numbers of escaping electrons. This is important for low-temperature plasmas such as those used in antihydrogen trapping. We perform a Monte Carlo simulation of our algorithm, and assess its robustness to intrinsic shot noise and external noise. Approximately $100$ particle counts are needed for an accuracy of $\pm 10 \%$ -- this provides a lower bound for measurable plasma temperatures of approximately $3\,\mathrm{K}$ for plasmas of length $1\,\mathrm{cm}$.}


\maketitle

\section{Introduction}\label{sec1}

\begin{figure}
  \centering
  \includegraphics[width=0.7\textwidth]{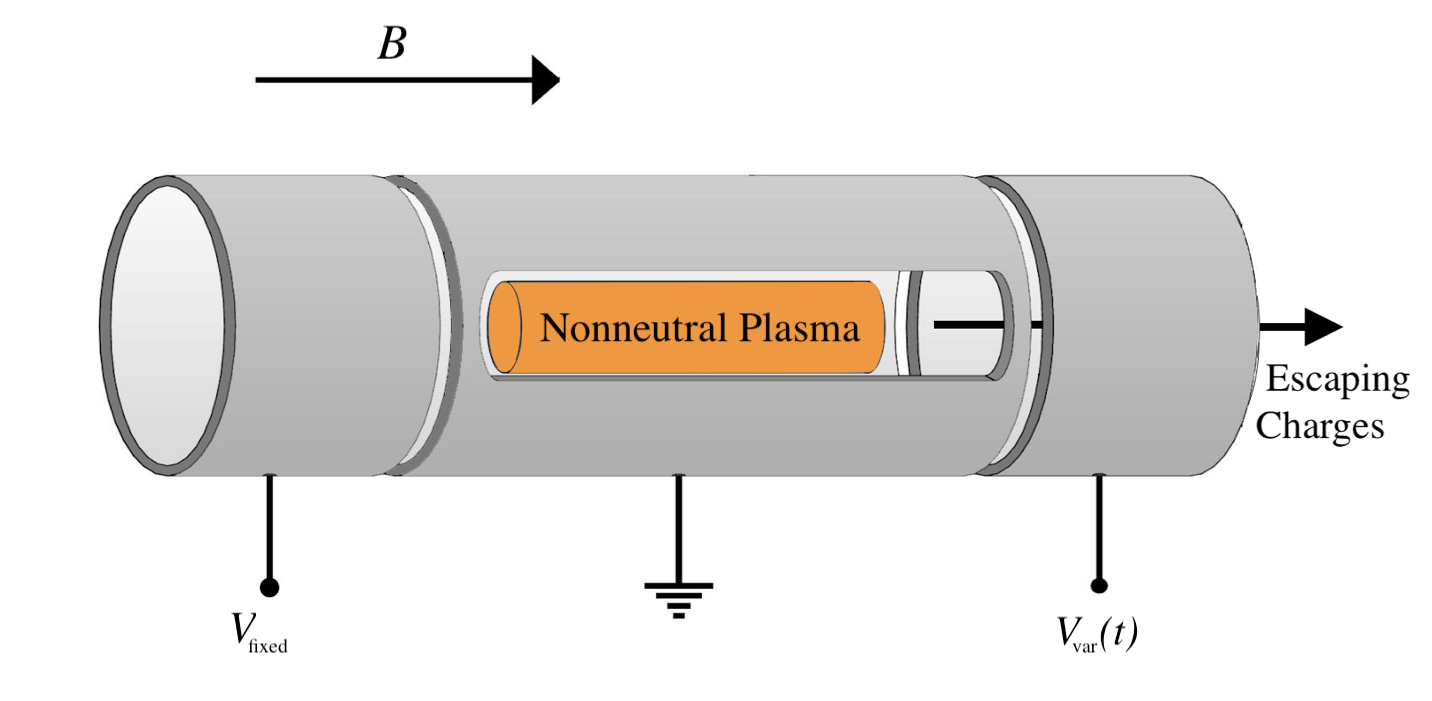}
  \caption{Schematic of the Penning-Malmberg trap apparatus. The cylindrical nonneutral plasma is confined radially by a magnetic field $B$ and axially by barrier electrodes with a fixed voltage $V_\mathrm{fixed}$ and a variable voltage $V_\mathrm{var}(t)$. In the temperature diagnostic, $V_\mathrm{var}(t)$ is lowered and individual charges with sufficiently high axial kinetic energy escape and are detected and logged by a silicon photomultiplier (not shown).}
  \label{fig:schematic}
\end{figure}

Nonneutral plasmas in Penning-Malmberg traps~\cite{MalmbergTrap} are confined radially by a strong axial magnetic field and are confined and manipulated longitudinally (along the magnetic field axis) by controlling electrode voltages on cylindrical segments of the trap, as illustrated in Fig.~\ref{fig:schematic}.

In principle, the trapped plasmas should cool to the temperature of their surroundings. In practice, however, this is typically not observed. It has been difficult  across a wide range of experiments to create plasmas colder than $100\,\mathrm{K}$ \cite{hunter2020plasma}. Nonetheless, constantly improving experimental techniques can generate plasmas with lower temperatures, approaching that of the electrodes \cite{baker2021crogenic,ams2022reduce}. With these improved experimental techniques, it is crucial to to be able to (1) robustly measure the plasma temperature, and (2) quantify the uncertainty underlying such a measurement. 

The most common experimental technique used to measure the plasma temperature is an evaporative protocol, first established by Hyatt, et al.~\cite{hyatt1987measurement} and further developed by Beck~\cite{beck1990thesis} and Eggleston, et al.~\cite{eggleston1992parallel}. In the protocol, one of the axially-confining electrode voltages $V_\mathrm{var}(t)$ is slowly reduced (see Fig.~\ref{fig:schematic}), allowing for the most energetic particles to escape. Particles escape times $\{t_i\}$ are recorded by a detector and the plasma temperature is then inferred from the barrier electrode voltage history.

The particle escape rate, assuming a Boltzmann distribution and restricting to times before enough particles escape and modify the self-potential of the plasma, matches closely with an exponential $|dN_\mathrm{esc}/d\Vb| \propto \exp(-q\Vb/k_\mathrm{B} T)$, which holds for $\Vb \gg k_\mathrm{B} T / q$. Here, $V_\mathrm{b}$ is the barrier voltage experienced by the plasma that is induced by the barrier electrode voltage $V_\mathrm{var}(t)$, $q$ is the particle charge, $k_\mathrm{B}$ is Boltzmann's constant, and $T$ is the plasma temperature. Eventually, enough plasma particles escape and the plasma self-potential changes, causing the escape rate to decrease and deviate from an exponential dependence on the voltage $\Vb$. Henceforth, we will refer to the former regime as the linear regime (as on a log plot it is a straight line), the latter as the saturated regime, and the transition between the two the bend-over regime (see Fig.~\ref{fig:fig1}). 

The standard method to infer $T$ fits a straight line fit on the log plot, thus, only using data from the linear regime. This corresponds, for typical parameters, to the escape of only $\sim$ 5\% of a Debye cylinder of plasma (a cylinder oriented along the trap magnetic field with a radius equal to the plasma's Debye length). For low temperature plasmas with $T < 5 \, \mathrm{K}$, this linear regime contains fewer than 50 particles, making for a intrinsically noisy fit. Moreover, the implementation of the linear-fit algorithm is complicated by the ambiguity of where the linear regime of the data ends, i.e., the point after which data should not be used in the fit, as well as by ambiguity in where the data begins, i.e., the point after which the data sufficiently exceeds the noise floor. Though there has been work trying to algorithmically detect these cutoff points in the escape data~\cite{evans2016thesis}, often times the cutoffs are manually decided, which introduces human error.

In order to incorporate data from the bend-over and saturated regimes, however, one must consider a model of the plasma and the protocol that includes these regimes. In earlier work by Beck~\cite{beck1990thesis},  a continuum model of the process was developed allowing for the calculation of a particle escape curve by solving a series of self-consistent ordinary differential equations (ODE) that describe the plasma at various moments of the protocol. 

In this paper, we implement a temperature-fitting algorithm that adapts Beck's model to single electron counts. Given binned escape data, our algorithm performs a maximum likelihood fit of plasma parameters, assuming that electron escapes follow an inhomogeneous Poisson process with escape rates determined by Beck's model. We quantify the robustness of the algorithm to intrinsic shot noise by running it on Monte Carlo simulation data. We find that for an estimator error of $10 \%$, around 100 particles are needed in combined linear and bend-over regimes; this corresponds to about a quarter of a Debye cylinder of plasma. From this, we infer a lower bound for measurable plasma temperature of around $3\,\mathrm{K}$ for a plasma of length $1\,\mathrm{cm}$, scaling inversely proportional to plasma length. This bound is about five times lower than what may be resolved using the straight-line fitter. Our temperature measurement method is shown to be robust to varying plasma parameters and external noise. 

Our work extends results obtained earlier~\cite{hunter2020plasma, shanman2016improved, evans2016thesis}, where silicon photomultiplier (SiPM) data which resolved  single particle escape times was analyzed with a Maximum Likelihood Estimator (MLE) fitter using only the standard ``straight line'' regime. 

The rest of this paper is as follows: In Sec. 2, we briefly review the temperature diagnostic protocol. In Sec. 3, we present Beck's model and our MLE fit for the temperature, and we compare our results to those obtained from a straight line fit. In Sec. 4, we describe the generation of Monte Carlo simulation data used to test the fitter, and in Sec. 5, we estimate using simulations the lowest possible measurable plasma temperature and the robustness of the fitter algorithm. In Sec. 6, we discuss our results and conclusions.

\section{Dynamic Evaporative Protocol} 
 
In the experimental protocol, the voltage $V_\mathrm{var}(t)$ of one of the axially confining barrier electrodes is slowly lowered, inducing a barrier voltage $\Vb$ (as measured at the axial center of the plasma $z = 0$). For a given barrier voltage $\Vb$, particles with energy

\begin{equation} \label{eq:escape-condition}
    E = \frac{mv_\parallel^2}{2} + q\phi(r, \theta) > q \Vb
\end{equation}
can cross the voltage barrier and escape. Here, $m$ is the particle's mass, $v_\parallel$ is its velocity in the axial direction, and $\phi$ is the electrostatic self-potential satisfying Poisson's equation

\begin{equation} \label{eq:poisson-equation}
    \nabla^2 \phi(\mathbf{x}) = - \frac{q n(\mathbf{x})}{\epsilon_0},
\end{equation}
where $n(\mathbf{x})$ is the number density of plasma particles, and $\epsilon_0$ is the vacuum permittivity. The boundary condition at the grounded wall at radius $R_\mathrm{w}$ is

\begin{equation} \label{eq:boundary-conditions}
    \phi(r = R_\mathrm{w}) = 0.
\end{equation}
Particles satisfying the inequality in Eq.~\eqref{eq:escape-condition} escape the plasma and are detected with a collector.

The diagnostic works in an intermediate timescale, wherein the (scaled) voltage reduction rate $\nu_\mathrm{protocol} = (k_\mathrm{B} T / q)^{-1} |d\Vb/dt|$ is faster than the electron collision frequency $\nu_{ee}$  and slower than then axial bounce frequency $\omega_z$: 

\begin{equation} \label{eq:timescales}
    \nu_{ee} \ll \nu_\mathrm{protocol} \ll \omega_z.
\end{equation}
In this case (i) the plasma does not rethermalize (i.e., does not repopulate higher-energy regions from which particles have already escaped) and (ii) a particle that has escaped and been detected at time $t_\mathrm{esc}$ can be assumed to have an energy well-approximated by $E_\mathrm{esc} = q \Vb(t_\mathrm{esc})$~\cite{eggleston1992parallel}.

Information about the distribution of axial energies (and, thus, the plasma temperature) is encoded in the escape times $\{t_i\}$ and the corresponding voltages $\{\Vb(t_i)\}$. However, this relationship is complex because  the potential energy term $q\phi$ in Eq.~\eqref{eq:escape-condition} cannot be ignored. First, for most plasmas the self-potential is significantly higher than thermal energy ($\phi \gg k_\mathrm{B} T$), varying on the order of the thermal voltage in the length scale specified by the Debye length: 

\begin{equation} \label{eq:lambda-debye}
    \lambda_\mathrm{D} := \sqrt{\frac{k_\mathrm{B} T \epsilon_0}{n_0 q^2}}
\end{equation}
(here $n_0$ is the initial plasma density at the radial center); and second, the self-potential is highest in the center axis, so that particles near the radial center escape earlier, thus altering the density profile of the plasma and thereby changing $\phi$.

\begin{figure}
\centering 
\includegraphics[width=1.0\textwidth]{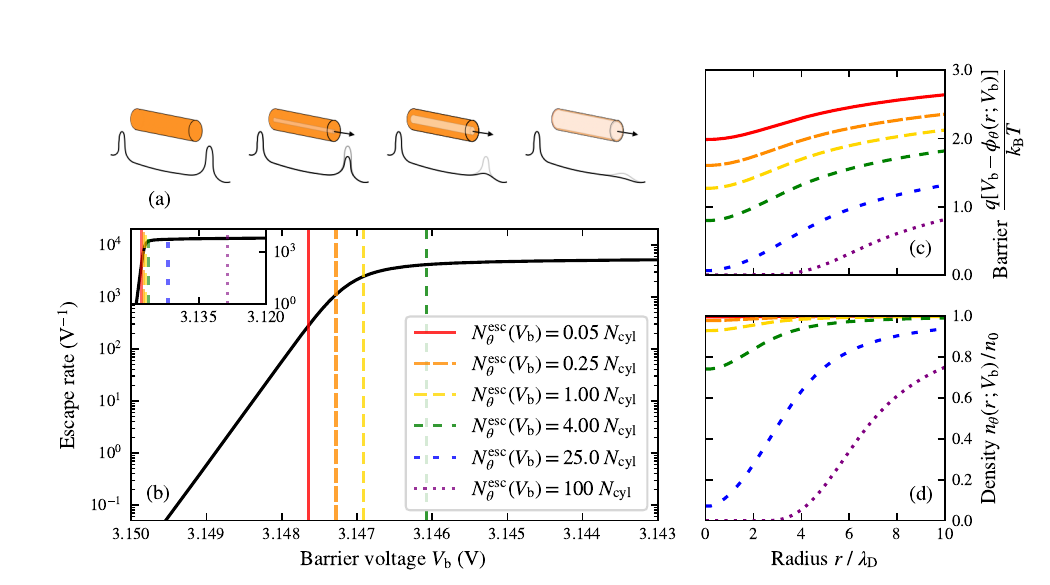}
\caption{(a) Cartoon schematic of the parallel temperature diagnostic protocol, from left to right. A cylindrical plasma (orange cylinder) is confined axially by voltage barriers (black curve below). One of the end barriers is lowered, and particles with high enough axial energy escape. Due to the self-potential, particles near the center are more likely to escape first. The escaped particles are detected and logged. (b) The escape rate $|d \Nesctheta / d\Vb|$ from the Beck model [solving Eqs.~\eqref{eq:poisson-eq-radial}--\eqref{eq:escape-curve-definition}] for a plasma with parameters $\boldtheta = (T = 26.6\,\mathrm{K}$, $n_0 = 10^8\,\mathrm{cm}^{-3}$, $R_\mathrm{p} = 1.0\,\mathrm{mm}$,  $\ell_\mathrm{p} = 1.0\,\mathrm{cm}$). Vertical lines represents times when a given fraction of a Debye cylinder has escaped. Initially the particles escape exponentially with reduced barrier voltage. However, as a more particles escape (from the vertical red line onward), the escape rate begins to saturate (see inset figure in (b)). (c) Effective barrier voltage (barrier voltage minus self-potential) and (d) local number density as a function of radius, at the corresponding moments of the protocol. The local barrier voltage is lower near the center of the plasma. Correspondingly, the particle density is lower near the center of the plasma, as more particles will have escaped from smaller $r$.}
\label{fig:fig1}
\end{figure}

Fig.~\ref{fig:fig1}(a) gives a schematic of the experimental protocol: as the barrier voltage $\Vb(t)$ is lowered, particles begin escaping from the center of the plasma where the self-potential is highest. These particles are detected by a collector, and the escape rate (from Beck's continuum-limit model discussed in the following section) is displayed in Fig.~\ref{fig:fig1}(b) in a log plot. For short times, the particle escape rate is roughly exponentially ($\log |dN_\mathrm{esc}/d\Vb| \approx q / k_\mathrm{B} T $); as more particles leave the plasma, the self-potential is modified and the escape rate saturates. This occurs roughly after 5\% of a Debye cylinder, $N_\mathrm{cyl} = n_0 \pi \lambda^2 \ell_\mathrm{p}$, escapes, and makes inferring the temperature from data using the escape voltages data $\{ V_i = \Vb(t_i) \} $ of particles in the bend-over and saturated regimes challenging. In order to infer a temperature estimate and error from the data, we combine Beck's plasma model with our maximum likelihood algorithm. 

\section{Fitting Algorithm}

\subsection{Beck's model} 

Here we briefly review Beck's continuum-limit model of the dynamic evaporative protocol \cite{beck1990thesis}. We assume that the plasma is initially a cylinder of radius $R_\mathrm{p}$, length $\ell_\mathrm{p}$, and uniform number density $n_0$, with $\ell_\mathrm{p} \gg R_\mathrm{p}\gtrsim 5 \lambda_D$. The plasma is thus determined by four parameters: 

\begin{equation} \label{eq:four-params-lambda}
    \boldtheta = (T, n_0, R_\mathrm{p}, \ell_\mathrm{p}).
\end{equation}
The initial total number of particles is then given by

\begin{equation}
    N_0 = n_0 \pi R_\mathrm{p}^2 \ell_\mathrm{p}, 
\end{equation}
and the number of particles in a Debye cylinder is 

\begin{equation}
    N_\mathrm{cyl} = n_0 \pi \lambda_\mathrm{D}^2  \ell_\mathrm{p} = 150.34 \ T \, (\mathrm{K}) \times \ell_\mathrm{p} \, (\mathrm{cm}).\label{eq:n-cyl-def}
\end{equation}
Note that $N_\mathrm{cyl}$ depends solely on the temperature $T$ and plasma length $\ell_\mathrm{p}$. $N_\mathrm{cyl}$ will be shown to be relevant for quantifying the robustness of the fitter to shot noise, as the number of particles in the linear and bend-over regimes is directly proportional to $N_\mathrm{cyl}$.

In the continuum limit, a deterministic escape curve for the cumulative number of escaped particles $\Nesctheta(\Vb)$ as a function of barrier voltage $\Vb$ may be constructed by solving a set of coupled Poisson-Boltzmann-like equations relating the number density of remaining (i.e., non-escaped) plasma $\ntheta(\mathbf{x}; \Vb)$ and the self-potential $\phitheta(\mathbf{x}; \Vb)$. Because of the azimuthal symmetry of the plasma and that $\ell_\mathrm{p} \gg R_\mathrm{p}$, only the radial dependence of $\ntheta$ and $\phitheta$ is relevant. Thus, Poisson's equation becomes a 2nd-order ODE in $r$:

\begin{equation}
  \frac{1}{r} \frac{d}{dr} \bigg( r \frac{d \phitheta (r; \Vb)}{dr} \bigg) = -\frac{q  \ntheta (r; \Vb)}{\epsilon_0}, \label{eq:poisson-eq-radial}
\end{equation} 
with the boundary conditions 

\begin{equation}
  \frac{d \phitheta (r = 0)}{dr} = 0 \quad\mathrm{and}\quad \phitheta(r = R_\mathrm{w}) = 0 \label{eq:poisson-init-conditions}
\end{equation}
determined by azimuthal symmetry and grounded trap walls, respectively. 

We also have from Eq.~\eqref{eq:escape-condition} and the assumption of a Maxwell-Boltzmann distribution a nonlinear $\Vb$-dependent relationship between particle density and self-potential energy 

\begin{equation} 
  \ntheta(r; \Vb) = \begin{dcases}
   n_0 \, \mathrm{erf} \sqrt{ \frac{q [ \Vb - \phitheta(r; \Vb) ] }{k_\mathrm{B} T } } \quad&\quad  \mathrm{for} \ r < R_p \\ 
  0  \quad&\quad \mathrm{otherwise}.
  \end{dcases} \label{eq:phi-n-relationship}
\end{equation} 
Here, the error function $\mathrm{erf}$ gives the fraction of the initial particles that do $not$ satisfy Eq.~\eqref{eq:escape-condition} and thus remain in the plasma.

Given a solution to Eqs.~\eqref{eq:poisson-eq-radial}--\eqref{eq:phi-n-relationship} for a given $\Vb$, the total number of remaining particles is 

\begin{equation} \label{eq:N-theta}
    N_\theta(\Vb) = 2 \pi \ell_\mathrm{p} \int_0^{R_\mathrm{p}} \ntheta(r; \Vb)  \, r \, dr,
\end{equation}
and the number of escaped particles is then given by 

\begin{equation} \label{eq:escape-curve-definition}
    \Nesctheta(\Vb) = N_0 - N_\theta(\Vb).
\end{equation}

Eqs.~\eqref{eq:poisson-eq-radial}--\eqref{eq:phi-n-relationship} may be numerically solved through leapfrog integration with a shooting method for mixed boundary conditions, and we provide details of our implementation in Appendix A. Further details and justifications of the model are contained in Beck's thesis~\cite{beck1990thesis}.

Fig.~\ref{fig:fig1}(b) shows the resulting escape rate curve $|d \Nesctheta(\Vb) / d \Vb|$ from the Beck model for typical plasma parameters $\boldtheta = (T = 26.6\,\mathrm{K}$, $n_0 = 10^8\,\mathrm{cm}^{-3}$, $R_\mathrm{p} = 1.0\,\mathrm{mm}$,  $\ell_\mathrm{p} = 1.0\,\mathrm{cm}$). The vertical lines in Fig.~\ref{fig:fig1}(b) correspond to the escape of different fractions of a Debye cylinder of plasma having escaped. Throughout a wide range of parameters, the linear regime cutoff roughly corresponds to $\Nesctheta(\Vb) = 0.05 N_\mathrm{cyl}$. Figs.~\ref{fig:fig1}(c) and ~\ref{fig:fig1}(d) plot the solutions for the local barrier height $q[\Vb - \phitheta(r; \Vb)]/k_\mathrm{B} T$  (the argument in the error function in Eq.~\eqref{eq:phi-n-relationship}), and the normalized density $\ntheta(r; \Vb) / n_0$ at values of $\Vb$ corresponding to the vertical lines in Fig.~\ref{fig:fig1}.(b). 

Varying the temperature changes both the slope of the linear regime of the curve (a lower temperature corresponds to a steeper slope) and the number of particles in $N_\mathrm{cyl}$ in Eq.~\eqref{eq:n-cyl-def}. Both of these effects affect the difficulty in resolving an accurate fit. On the other hand, varying the plasma length $\ell_\mathrm{p}$ affects only the number of particles in $N_\mathrm{cyl}$, with longer plasmas having a greater number of particles in a Debye cylinder. Generally speaking, the parameters $n_0$ and $R_\mathrm{p}$ do not affect the shape of the curve very much as long as the ratio between plasma radius and Debye length is much greater than one ($x_\mathrm{p} = R_\mathrm{p} / \lambda_\mathrm{D} \gg 1$), which is the case for typical plasma parameters of interest here (where $x_\mathrm{p} \sim 100$);  varying $n_0$ and $R_\mathrm{p}$ with fixed $T, N_\mathrm{cyl}$ offsets the curve horizontally, changing the numerical values of $\Vb$ that correspond to the linear regime and the bend-over regime.

\subsection{Maximum Likelihood Estimation of Plasma Parameters}
\label{section:likelihood-maximization}

The idea of exploiting a MLE for the temperature analysis of the temperature diagnostic was first developed by Evans~\cite{evans2016thesis} in a study of single-particle-resolution SiPM data \cite{hunter2020plasma, shanman2016improved}.

In experiments, escape voltages $\{\Vb(t_i) \, | \, \mathrm{particle \ escape \ at \ } t_i\}$ are obtained as binned data $\mathcal{D} = (\mathcal{N}, \mathcal{B}) $, with $N_k \in \mathcal{N}$ counts observed within $V_k, V_{k+1} \in \mathcal{B}$ barrier voltage bin limits (i.e., $V_k > \Vb > V_{k+1}$) for each bin number $k = 1, 2, ..., K$. (Here, we have assumed the convention $V_k > V_{k+1}$, as barrier voltages are lower further along the protocol.) Under the assumption that discrete particle arrivals follow a non-homogeneous Poisson process with the expected cumulative escapes equal to $\Nesctheta(\Vb)$, the likelihood of observing binned data $\mathcal{D}$ for plasma parameters $\theta$ is

\begin{equation} \label{eq:likelihood}
    L(\boldtheta; \mathcal{D}) = \prod_{k} e^{-\mu_{k}(\boldtheta)} \frac{\mu_k(\boldtheta)^{N_k}}{N_k!},
\end{equation}
where $\mu_{k}(\boldtheta)$ is the expected number of counts occurring between $\Vb = V_{k}$ and $\Vb = V_{k+1}$, given by their difference in the cumulative escape curve

\begin{equation} \label{eq:expected-counts}
    \mu_{k}(\boldtheta) = N_\theta^\mathrm{esc}(V_{k + 1}) -  N_\theta^\mathrm{esc}(V_{k}). 
\end{equation}

The MLE for the plasma parameters is the argmax of the likelihood Eq.~\eqref{eq:likelihood}; equivalently, it is the argmin of the negative log likelihood

\begin{equation} \label{eq:MLE-negative-log-likelihood}
   \hat{\boldtheta} = \operatorname*{argmin}_{\boldtheta} \, (-\log L (\boldtheta; \mathcal{D})) = \operatorname*{argmin}_{\boldtheta} \sum_{k} \bigg( \mu_{k}(\boldtheta) - N_k   \log {\mu_k(\boldtheta)} + \log{N_k!} \bigg).
\end{equation}
In order to evaluate the argument of Eq.~\eqref{eq:MLE-negative-log-likelihood} for a particular $\boldtheta$, the escape curve must be evaluated at each of the voltage bin limits $V_k \in \mathcal{B}$ to obtain the expected counts in Eq.~\eqref{eq:expected-counts}. This requires solving Eqs.~\eqref{eq:poisson-eq-radial}--\eqref{eq:escape-curve-definition} for each bin limit barrier voltage $\Vb = V_k$. The negative log likelihood function is run through a minimization algorithm---in our implementation we use the Nelder-Mead algorithm \cite{nelder-mead-orig}---to obtain the fit $\hat{\theta}$.

\begin{figure}
\centering
    \includegraphics[width=.7\textwidth]{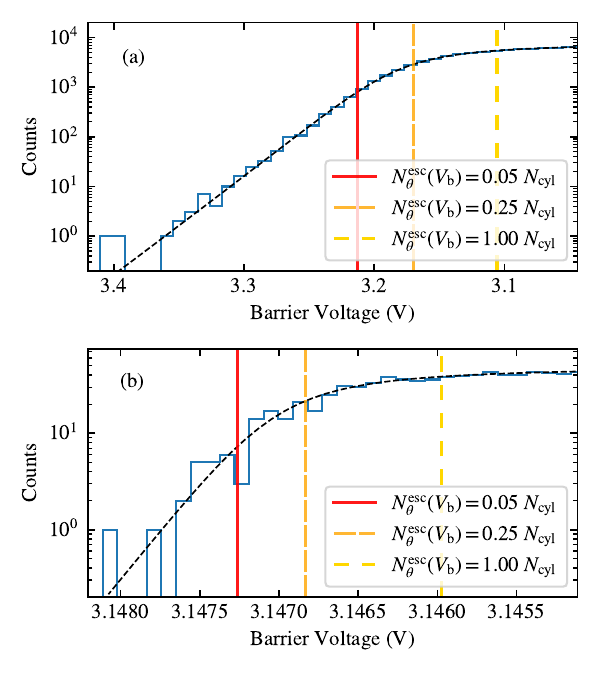}
 \caption{Binned simulation data (blue histogram) and the computed fit (dashed black line) for plasmas with parameters (a) $\boldtheta = (T = 266\,\mathrm{K}$, $n_0 = 10^8\,\mathrm{cm}^{-3}$, $R_\mathrm{p} = 1.0\,\mathrm{mm}$,  $\ell_\mathrm{p} = 1.0\,\mathrm{cm}$) and (b) same parameters as in (a) except the temperature is now $T = 2.66\,\mathrm{K}$ . For ease of visualization, histogram bins contain multiple counts. Fits were made using single-count resolution histograms and the first $\Ndata = 0.25 \ N_\mathrm{cyl}$ escaping particles (100 and 10000 particles, respectively), i.e., using bins to the left of the dashed vertical orange line. Beyond this regime, diocotron instabilities may cause changes to the distribution function not captured by the model and lead to chaotic particle escapes. The best fit temperature parameters are $\hat{T} = 266.42\,\mathrm{K}$ and $\hat{T} = 2.67\,\mathrm{K}$ respectively.}
\label{fig:fig2}
\end{figure}

Though Beck's model requires four input parameters $\theta$, in practice the plasma length $\ell_\mathrm{p}$ and total charge $Q = q N_0$ are easily experimentally accessible quantities, and their knowledge may be used to reduce the dimension of parameter space Eq.~\eqref{eq:four-params-lambda} to two through the equality constraints

\begin{equation} \label{eq:constrained-parameter-space}
  \boldtheta \in  \{ (T, n_0, R_\mathrm{p}, \ell_\mathrm{p}) \ | \ \ell_\mathrm{p} = \ell_\mathrm{p}^\mathrm{true}, \, n_0 \pi R_\mathrm{p}^2 \ell_\mathrm{p}= N_0^\mathrm{true}\}.
\end{equation}
This reduction of dimensionality aids in the numerical convergence of the temperature fitter, as occasionally we have observed a parameter redundancy (i.e., for some Monte Carlo data $\mathcal{D}$ there were different parameters in some set $\theta \in \Theta_{\mathrm{redundant}}$ that returned the same likelihood $L(\theta; \mathcal{D})$ (see Appendix C)). Fortunately, in these cases of parameter redundancy, the estimated temperature parameter $\hat{T}$ varied by approximately $\sim 2\%$. This does not significantly degrade the accuracy of the estimated temperature, and is easily avoided through introducing the above constraints using known values of the plasma length and particle number.

In what follows, we assume that the initial values of the length $\ell_\mathrm{p}$ and total charge $Q $ are known precisely.  A more complex model, not pursued here, could include errors in these quantities by assuming measurement priors. Further, $\ell_\mathrm{p}$ is assumed to be constant with time.  This is true only so long as few particles have escaped and $\{\Vb(t_i)\}$ has not changed significantly.

\subsection{Comparison with a straight line fit}

In practice a ``straight line fitter'' is commonly used to determine the parameters temperature $T$ and rate amplitude $A$. The fitter is constructed based on the observation that the escape energy distribution near the beginning of the protocol may be approximated with a Boltzmann distribution: $N^\mathrm{esc}(\Vb) \approx A \exp(-q\Vb / k_\mathrm{B} T)$. 

As developed in Ref.~\cite{evans2016thesis}, the ``straight line fitter'' is a maximum likelihood estimator for parameters $(T, A)$ that minimizes Eq.~\eqref{eq:MLE-negative-log-likelihood}, but with expected counts given by

\begin{equation}
  \mu_k^{\mathrm{sl}} (A, T) = A \exp \bigg( - \frac{q V_{k}}{k_\mathrm{B} T} \bigg) \bigg[ \exp \bigg(\frac{q \Delta V_k}{k_\mathrm{B} T} \bigg) - 1 \bigg]
\end{equation}
where $\Delta V_k = V_k - V_{k+1}$ is the bin width. In the case that all bins have the same bin width $\Delta V$, this gives a linear relationship between the log expected counts and barrier voltage ($\log \mu(\Vb) = - (q / k_\mathrm{B} T) \Vb + \mathrm{const}$), with the slope given by the inverse temperature. 

The approximation is surprisingly good, in particular in the high-temperature limit \cite{beck1990thesis,eggleston1992parallel,hyatt1987measurement}, and its justification can be found in \cite{beck1990thesis}. However, the straight-line assumption is limited to the region where the curve is linear, which we have found to include only the first $0.05N_\mathrm{cyl}$ of escaping plasma. Escape data that could provide additional temperature information is thereby discarded, leading to an unnecessarily greater uncertainty in the temperature estimate. 
Furthermore, it may be difficult to decide where the linear regime ends, i.e., where to cutoff the data to be used in the straight line fitter. Though work has been done to attempt to automatically determine this cutoff \cite{evans2016thesis}, it is often determined manually, introducing human error. 

\section{Numerical Modeling} \label{section:simulation}

In order to quantify the performance of our maximum likelihood fitting algorithm, we compare the statistics of the MLE estimated parameters $\hat{\theta}$ for Monte Carlo simulation particle escape data to their ground truth values $\boldtheta^\mathrm{true}$ used in generating the Monte Carlo data. 

\subsection{Monte Carlo simulations} 

The Monte Carlo simulation is consistent with the Beck Model pre-continuum limit. In brief, for parameters $\boldtheta$ particles were randomly initialized with a radial position $r$ drawn with probability $p(r) \propto r$ with $r \in (0, R_\mathrm{p})$; and a velocity $v$ is drawn from a Maxwellian with temperature $T$ with probability $p(v) \propto \exp(-mv^2/2k_\mathrm{B}T)$.

We make a key assumption that the plasma always maintains a azimuthally symmetric density. This is critical for computational tractability. In reality, particles are discrete when they hit the diagnostic. The model of the plasma assumes that in the calculation of the space-charge potential each charge can be  smeared   onto a cylindrical shell  of uniform charge. 
The radial self-potential $\phi(r)$ is then calculated by solving Poisson's equation with $\phi(R_\mathrm{w}) = 0$ to find the total energy of each particle $E_i = mv_i^2/2 + q \phi(r_i)$. 

After the generation of the initial plasma,  the particle $i$ with the highest total energy is removed, with its escape voltage $V_i = E_i / q$ recorded. The self-potential energy $\phi(r_i)$ is then recalculated for each remaining particle, and the total energy $E_i$ updated. This is repeated until all particles escape to produce the voltages $\{V_i\}$ for a single plasma and protocol instance (see Fig.~\ref{fig:fig2}). We provide more details of our Monte Carlo simulations in the Appendix B.

Examples of simulation data and best-fit curves are plotted in Fig.~\ref{fig:fig2}(a)-(b) for two plasma with different temperatures $T = 266\,\mathrm{K}$, and $T = 2.66\,\mathrm{K}$, as well as the MLE escape curve fit from Beck's model. A few notable points are that (1) for the lower temperature plasma~\ref{fig:fig2}(b) there is a significantly smaller range in the barrier voltages for the particle escapes, and (2) for the lower temperature there are fewer total particles in the linear regime (to left of the orange line in~\ref{fig:fig2}(b)). Fewer particles creates a noisier fit, as seen in~\ref{fig:fig2}(b). As found by Eggleston~\cite{eggleston1992parallel}, the straight line region corresponds to only around one decade of usable escape data for temperatures less than 100 Kelvin.

After parameters are estimated for each sample, the relative bias and the relative standard error (i.e., square root of estimator variance) of the temperature estimator are given by 

\begin{equation}
    b = \frac{\langle \hat{T} \rangle - T}{T} \quad\mathrm{and}\quad \varepsilon = \frac{\sqrt{\langle \hat{T}^2 \rangle - \langle \hat{T} \rangle^2 }}{T}, 
\end{equation}
where the brackets denote the average over the ensemble samples $\langle f(\hat{T})\rangle := S^{-1} \sum_{s = 1}^S f(\hat{T}_s) $. 

 \section{Simulation Results} \label{section:results} 

\subsection{Minimum temperature limit}
\begin{figure}
\centering
    \includegraphics[width=1.0\textwidth]{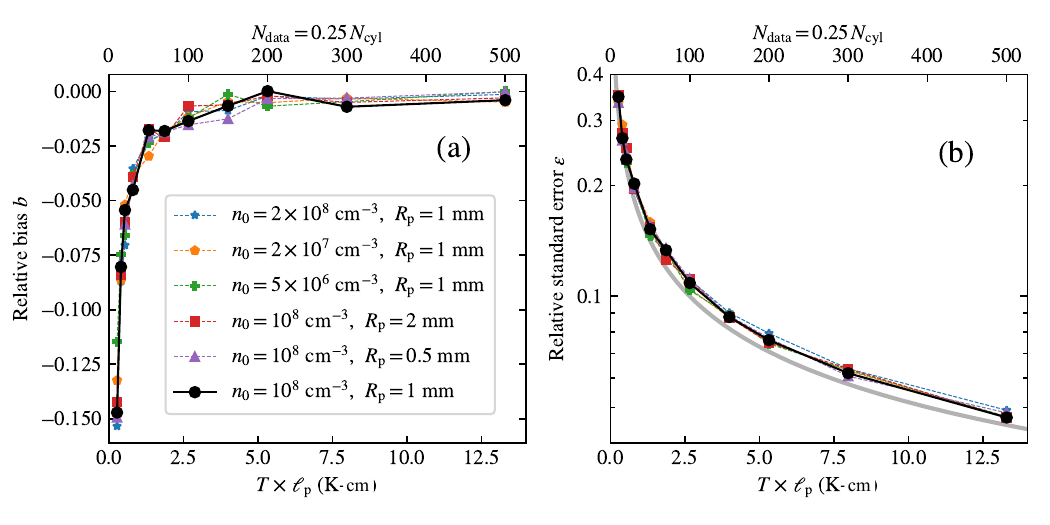}
\caption{ Numerically obtained relative bias (a) and relative standard error (b) of our estimator for different plasma parameters. Plasma length is set to $\ell_\mathrm{p} = 1\,\mathrm{cm}$, temperature $T$ is specified by the horizontal axis (and subsequently the number of used particles $\Ndata = 0.25 N_\mathrm{cyl} \propto \ell_\mathrm{p} T$), and different curves are for varying $R_\mathrm{p}$ and $n_0$ as specified in the legend. There are no significant differences in the quantitative behavior of our temperature diagnostic across data sets of plasmas with differing radii and densities. The transparent curved black line in the right plot is a reference $\varepsilon = 1 / \sqrt{\Ndata}$. The empirical fitting errors are just slightly higher than the black line. An error of 10\% corresponds to a $T \times \ell_\mathrm{p}$ product of around $3\,\mathrm{K\mbox{--}cm}$.
}
\label{fig:fig3}
\end{figure}

For our baseline simulations, we consider plasmas with non-temperature parameters set to typically observed values of $n_0 = 10^8 \,\mathrm{cm}^{-3}, \ R_\mathrm{p} = 1.0\,\mathrm{mm}, \ \ell_\mathrm{p} = 1.0 \,\mathrm{cm}$; and varying temperatures $T$ between $0.266 \,\mathrm{K}$ and $13.3\,\mathrm{K}$, corresponding to the number of particles within one Debye cylinder $N_\mathrm{cyl}$ to vary between $40$ and $2000$. Data is binned with single-particle resolution, i.e., each voltage bin has either zero or a single count. 

In practice, as more particles escape, the chance of observing diocotron instabilities increases, limiting the amount of escape data that are actually usable. Because it is experimentally difficult to determine when the diocotron instability sets in, to be conservative we have chosen to limit our fitter to using only the first $\Ndata = 0.25 \, N_\mathrm{cyl}$ escapes (i.e., $\{V_i^{(s)} \, | \, i \leq 0.25 N_\mathrm{cyl} \}$). This corresponds to the linear regime and, in addition, some of the bend-over regime (cf., to the left of the orange line Fig.~\ref{fig:fig1}(b)). 

The error for these baseline simulations is depicted in the black bold line in Fig.~\ref{fig:fig3}. The error as a function of number of particles is slightly higher than $1/\sqrt{\Ndata}$, the latter plotted in a black dashed line as a reference. 

A main result of this section is that to obtain an error of $\varepsilon \leq 10\%$, we need a plasma that yields slightly more than $\Ndata = 100$ counts within a quarter Debye cylinder. This corresponds to $T \ell_\mathrm{p} = 3\,\mathrm{K\mbox{--}cm}$. In other words, for a plasma with $\ell_\mathrm{p} = 1 \, \mathrm{cm}$, a temperature of $T = 3\,\mathrm{K}$ is the lowest we can measure if we desire a 10\% accuracy of temperature fit. 

We find a slight negative bias $b$ for small $T \ell_\mathrm{p}$; the bias  magnitude is eclipsed by the standard error of the estimator (e.g., at $T \ell_\mathrm{p} = 3\,\mathrm{K\mbox{--}cm}$ the error $\varepsilon = 10\%$ the relative bias is around $2\%$).

\subsection{Robustness to variations in plasma density and radius}

We simulated ensembles of plasmas with differing values of $n_0$ and $R_\mathrm{p}$. The results are shown in Fig.~\ref{fig:fig3}. As explained in the paragraph before Section~\ref{section:likelihood-maximization}, differing values of $n_0$ and $R_\mathrm{p}$ mostly change the value of $\Vb$ when escapes begin. We do not see significant quantitative differences in the shape of the error curves, as seen in Fig.~\ref{fig:fig3}, illustrating the robustness of our fitting algorithm to these variations of $n_0$ and $R_\mathrm{p}$.

\subsection{Finite bin width}

Given continuous hit data generated from Monte Carlo simulations $\{ V(t_i), \}$, the histogram bin size $\Delta V$ used in data binning is a hyperparameter (though in practice it is limited by the product of the temporal resolution of the detector $\Delta t$ times the voltage ramp speed limited by Eq.~\eqref{eq:timescales}, i.e., $\Delta V \gtrsim | d\Vb/dt| \Delta t$). A larger value of $\Delta V$ decreases computation time (i.e., the number of times $\Nesctheta(\Vb)$ needs to be evaluated per iteration), but introduces a binning error to the fit. We find numerical convergence for $\Delta V \leq 20 \, \delta V_\mathrm{min}$ where $\delta V_\mathrm{min} = (q / 2\pi \ell_\mathrm{p} \epsilon_0) \ln (R_\mathrm{w}/ R_\mathrm{p})$ is the minimum difference in escape voltages between adjacent escapes in our Monte Carlo model. We observe a deviation of $1.1\%$ in the root mean square when using the histogram bin size $\Delta V = 20 \, \delta V_\mathrm{min}$, compared to using $\Delta V = 0.1 \, \delta V_\mathrm{min}$ across the same population.

\subsection{Inclusion of data beyond the linear regime}

\begin{figure}
\centering
    \includegraphics[width=1.0\textwidth]{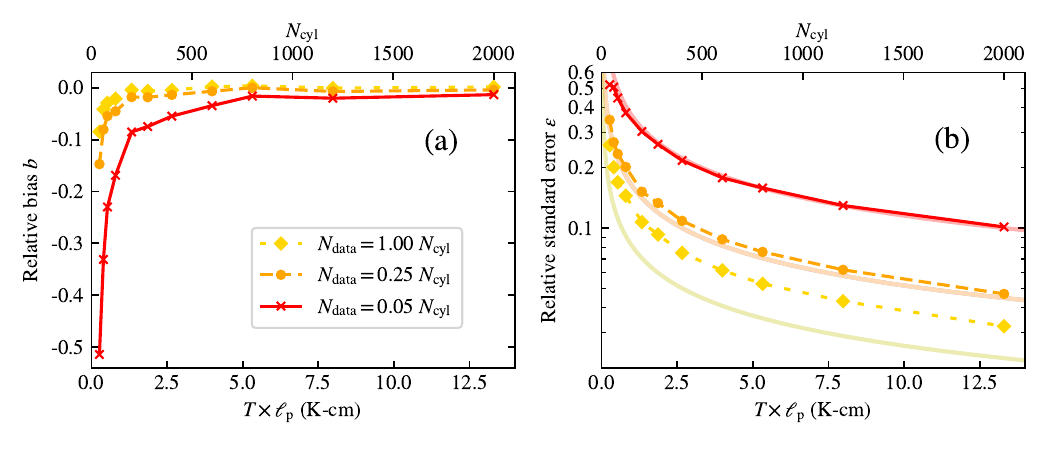}
\caption{Numerically obtained relative bias (a) and relative standard error (b) for our temperature estimator using a different amount of $\Ndata$ in the fitter (cf., Fig.~\ref{fig:fig1}.(b)). Red represents the linear regime  (5\% of a Debye cylinder), orange includes some data in the bend-over regime  (25\%  of a Debye cylinder, the same curves are illustrated as the black-dotted curves for $(n_0 = 10^8 \ \mathrm{cm}^{-3}, R_\mathrm{p} = 1 \ \mathrm{mm})$ in Fig.~\ref{fig:fig3}), and yellow includes a full Debye cylinder of simulated escape data.   Solid lines on right plot are $\varepsilon = 1 / \sqrt{\Ndata}$, as a reference. The linear regime matches  closely to the theoretical expectation of ($1 / \sqrt{\Ndata}$) dependence, while inclusion of a larger fraction of the plasma, e.g., a full Debye cylinder, does not scale as favorably, as seen by the yellow  simulation results, which lie above the $\varepsilon = 1 / \sqrt{\Ndata}$ solid yellow line.} \label{fig:fig4}
\end{figure}

Here, we vary the data in the fitter by considering the linear regime $\Ndata = 0.05 \, N_\mathrm{cyl}$, and the saturated regime $\Ndata = 1.00 \, N_\mathrm{cyl}$ (see Fig.~\ref{fig:fig1}(b)). Fig.~\ref{fig:fig4} shows that using only the linear regime data (yellow curves) provides poorer fits (an error of 10\% is reached at $T \ell_\mathrm{p} \approx 13\,\mathrm{K\mbox{--}cm}$), while using a full Debye cylinder of plasma for the fit provides for a more accurate fit (an error of 10\% is reached at  $T \ell_\mathrm{p} \approx 1.5\,\mathrm{K\mbox{--}cm}$). Plotted as well are $1 / \sqrt{\Ndata}$ curves, the theoretical expectation for fitting an exponential escape curve to exponential escape data. We see that the $\Ndata = 0.05 \, N_\mathrm{cyl}$ case matches nearly exactly with its theoretical curve (except for the uppermost data-points, which have $\Ndata = 2$ and $\Ndata = 5$ particles respectively,  trials are likely over-fit), while for $\Ndata = 1.00 \, N_\mathrm{cyl}$ the scaling is not as favorable.
This may be due to the fact that beyond the linear regime, each additional particle used in the temperature diagnostic contributes to a smaller reduction of error than would have been the case if the escape rate did not saturate and continued to be exponential.

The implementation of the  temperature diagnostic using data beyond the linear regime is desirable and can obtain a reduced error in the temperature estimate. Other effects, such as diocotron instabilities, which are beyond the scope of this analysis, may limit performance as more particles are extracted from the plasma.

\subsection{External noise}
We model the effect of external noise with a homogeneous Poisson process with a constant rate $\lambda_{\mathrm{ext}}$ per change in voltage. The expected counts per bin (c.f., Eq.~\eqref{eq:expected-counts}) 
 under our model is now

\begin{equation} \label{eq:cumulative-counts-with-noise}
    \mu_{k}(\boldtheta, \lambda_\mathrm{ext}) = \Nesctheta(V_{k + 1}) - \Nesctheta(V_k) + \lambda_\mathrm{ext}| V_{k+1} - V_{k}|,
\end{equation}
as the expected external noise counts in bin $(V_k, V_{k+1})$ is $\mu_{k, \mathrm{ext}} = \lambda_{\mathrm{ext}} |V_{k+1} - V_k|$, and the sum of two independent Poisson distributions is a Poisson distribution with a summed expected counts parameter.

\begin{figure}
\centering
    \includegraphics[width=1.0\textwidth]{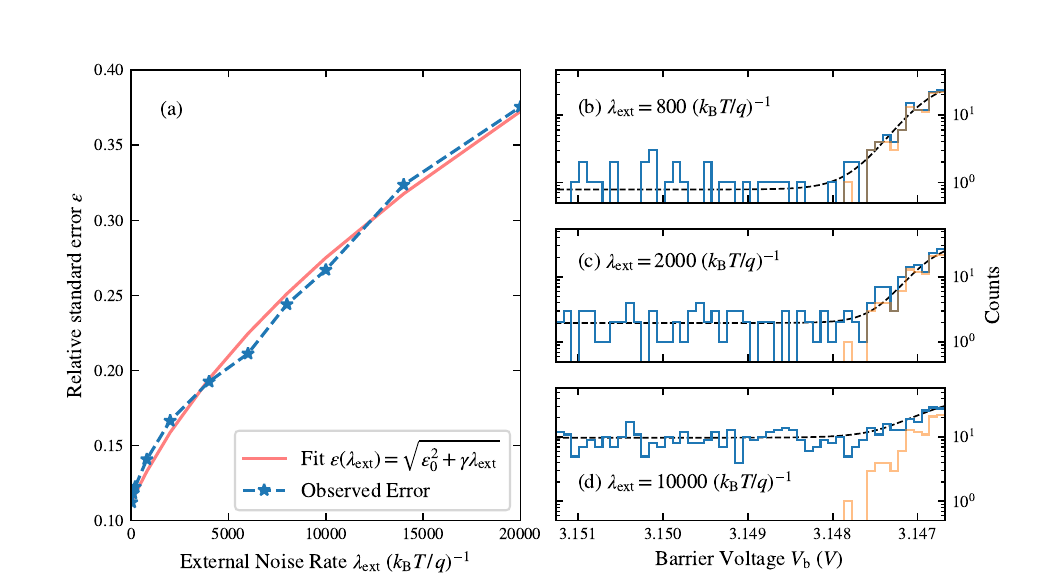}
\caption{(a) Estimator relative standard error for a plasma with parameters $(T = 2.66\,\mathrm{K}$, $n_0 = 10^8\,\mathrm{cm}^{-3}$, $R_\mathrm{p} = 1.0\,\mathrm{mm}$,  $\ell_\mathrm{p} = 1.0\,\mathrm{cm}$), as a function of added external homogeneous Poisson noise. $\Ndata = 0.25 N_\mathrm{cyl}$ particles are used in the fit, corresponding to $100$ particles. The likelihood function used in fitting routine is augmented to include a term for the external noise for each bin (see Eq.~\eqref{eq:cumulative-counts-with-noise}). The observed total error matches well with the sums-in-quadrature error $\varepsilon(\lambda_\mathrm{ext}) = \sqrt{\varepsilon^2 + \gamma \lambda_{\mathrm{ext}}}$ (unbroken red line). Here $\varepsilon_\mathrm{ext}^2 = \gamma \lambda_{\mathrm{ext}}$ corresponds to the squared amplitude of a homogeneous Poisson process, with the best fit $\gamma = 6.31 \times 10^{-6} k_\mathrm{B}T/q$. (b-d) the  simulated escape data (orange) with varying amounts of external noise; the blue line is the sum of the signal and external noise and the dashed black line is the fit (Eq.~\eqref{eq:cumulative-counts-with-noise}). (b) For  $\lambda_\mathrm{ext} = 8 \times 10^2 \ q/k_\mathrm{B}T$, the fitter gives temperature $\hat{T} = 2.830\,\mathrm{K}$. (c)  For $\lambda_\mathrm{ext} = 2 \times 10^3 \ q/k_\mathrm{B}T$, the fitter gives temperature $\hat{T} = 2.468\,\mathrm{K}$. (d)  For $\lambda_\mathrm{ext} = 10^4  \ q/k_\mathrm{B}T$, the fitter gives temperature $\hat{T}=3.438\,\mathrm{K}$.
}
\label{fig:fig5}
\end{figure}

In principle, the estimator errors from intrinsic shot noise and homogeneous external noise should sum in quadrature as

\begin{equation} \label{eq:error-quadrature}
    \varepsilon^2_\mathrm{total} = \varepsilon_0^2 + \varepsilon^2_\mathrm{ext},
\end{equation}
where $\varepsilon_0$ is the standard error without external noise.
We ran our simulation for a single set of parameters $\boldtheta = (T = 2.66\,\mathrm{K}$, $n_0 = 10^8\,\mathrm{cm}^{-3}$, $R_\mathrm{p} = 1.0\,\mathrm{mm,}$  $\ell_\mathrm{p} = 1.0\,\mathrm{cm}$), with varying amounts of external noise added. As before, we use only $\Ndata = 0.25 N_\mathrm{cyl}$, and we see that the observed error as a function of noise amplitude, plotted in Fig.~\ref{fig:fig5}, shows a good agreement with Eq.~\eqref{eq:error-quadrature}.

Experimentally, the noise can be essentially zero with an advanced microchannel plate (MCP) and silicon photomultiplier (SiPM) system, but can be much larger with less advanced systems \cite{hunter2020plasma}.

\section{Conclusion} 

In this paper we have shown that shot noise imposes a lower limit of about $3 \,\mathrm{K}$ for a  plasma of length $1 \,\mathrm{cm}$,  on the  nonneutral plasma temperatures that can be measured with the standard MCP and SiPM diagnostic.   The data analysis required a new algorithm which includes data that has been previously neglected (i.e., electrons that arrived after the end of the linear regime of the diagnostic). This limit is seen to be inversely proportional to the length of the plasma and for typical parameters independent of plasma density. The algorithm has been demonstrated numerically on synthetic data and is ready for experimental realization. Many effects are not considered here, such as the variation of the plasma length with barrier voltage as the plasma escapes and variations of the barrier voltage with radius. Since these errors may cause systematic biases in the measured temperatures, they are deserving of further exploration.

\subsection*{Declarations}

\subsubsection*{Availability of data and materials}

All data generated and analysed during this study are available from the corresponding author on reasonable request.

\subsubsection*{Competing interests}

The authors declare that they have no competing interests.

\subsubsection*{Funding}

AZ was supported by the Department of Defense (DoD) through the National Defense Science \& Engineering Graduate (NDSEG) Fellowship Program. This work was supported by the DOE OFES and NSF-DOE Program in Basic Plasma Science. 

\subsubsection*{Authors' contributions}

JF developed the concept of utilizing data beyond the linear regime. All authors contributed to the algorithm design. Data generation and analysis were performed by AZ. The first draft of the manuscript was written by AZ and all authors commented on previous versions of the manuscript. All authors read and approved the final manuscript.

\subsubsection*{Acknowledgements}

The authors thank Malcolm Lazarow, Andrew Charman, Tyler Maltba, and Eugene Kur for useful discussions. We benefited from preliminary work on Poisson-based temperature fits by Sabrina Shanman and Lena Evans. Simulations were principally performed on the Savio computational cluster resource, provided by the Berkeley Research Computing program at the University of California, Berkeley which is supported by the UC Berkeley Chancellor, Vice Chancellor for Research, and Chief Information Officer).

\bibliography{sn-bibliography}

\appendix

\newpage 

\section{Numerical Methods} 

In this Appendix, we present our numerical implementation for solving Beck's model to obtain $N_\mathrm{esc}(\theta)$ used in our temperature fitter.

\subsection{Derivation of the model}

Inserting Eq.~\eqref{eq:phi-n-relationship} into  Eq.~\eqref{eq:poisson-eq-radial} yields a cylindrical Poisson-Boltzmann-like  equation,  (cf., \cite{poisson-boltzmann}) for $\phitheta$ in the region  $r \in [0, R_\mathrm{p}]:$

\begin{equation} \label{eq:poisson-boltzmann}
    \frac{1}{r} \frac{d}{dr} \bigg( r \frac{d \phitheta (r; \Vb)}{dr} \bigg) = -\bigg( \frac{q  n_0 }{\epsilon_0} \bigg) \mathrm{erf} \sqrt{ \frac{q ( \Vb - \phitheta(r; \Vb) ) }{k_\mathrm{B} T } }.
\end{equation}

Equation~\eqref{eq:poisson-boltzmann} with the boundary condition $\phitheta(r = R_\mathrm{w}) = 0$ from $r = R_\mathrm{w}$ to $r = R_\mathrm{p}$ can be integrated to yield

\begin{equation} \label{eq:boundary-plasma-radius}
    \phitheta(R_\mathrm{p}; \Vb) = \frac{q N_\theta(\Vb)}{2 \pi \epsilon_0 \ell_\mathrm{p} } \ln \bigg(\frac{R_\mathrm{w}}{R_\mathrm{p}} \bigg),
\end{equation}
where $qN_\theta(\Vb) / \ell_\mathrm{p}$ serves as the charge per unit length. This is valid for an infinite length line charge limit (i.e., large $\ell_\mathrm{p}$) with azimuthal symmetry, which is assumed.

This may be further simplified by expressing

\begin{align} \label{eq:cumulative-definition-new}
  N_\theta(\Vb) &= 2\pi \ell_\mathrm{p} \int_0^{R_\mathrm{p}} \ntheta(r; \Vb) \, r dr \nonumber \\
  &=  2\pi \ell_\mathrm{p} \int_0^{R_\mathrm{p}} n_0 \, \mathrm{erf} \bigg(\sqrt{\frac{q(\Vb - \phitheta(r; \Vb))}{k_\mathrm{B} T }}\bigg)  r dr \nonumber \\ 
  &= - \bigg( \frac{2\pi \ell_\mathrm{p} \epsilon_0}{q} \bigg) \int_0^{R_\mathrm{p}} \frac{1}{r} \frac{d}{dr} \bigg( r \frac{d\phitheta(r; \Vb)}{dr} \bigg)  r dr \nonumber \\
  &= -\bigg( \frac{2\pi \ell_\mathrm{p} \epsilon_0 R_\mathrm{p}}{q} \bigg) \frac{d\phitheta(R_\mathrm{p}; \Vb)}{dr}.
\end{align}
In the third line we insert  Eq.~\eqref{eq:poisson-boltzmann}, and in the fourth line we take the integral in $r$. 

Inserting this expression for $N(\Vb)$ into Eq.~\eqref{eq:boundary-plasma-radius} yields the Robin boundary condition 

\begin{equation} \label{eq:boundary-plasma-radius2}
  \phitheta(R_\mathrm{p}; \Vb) + R_\mathrm{p} \ln \bigg( \frac{R_\mathrm{w}}{R_\mathrm{p}} \bigg)\frac{d\phitheta(R_\mathrm{p}; \Vb)}{dr} = 0.
\end{equation}
Our system of equations is now entirely in the single function $\phitheta(r; \Vb)$ over the domain $r \in [0, R_\mathrm{p}]$, and consists of the ODE Eq.~\eqref{eq:poisson-boltzmann}, the boundary condition Eq.~\eqref{eq:boundary-plasma-radius2}, and the Neumann boundary condition at the origin

\begin{equation} \label{eq:boundary-condition-origin-repeat}
  \frac{d\phitheta(0; \Vb)}{dr} = 0.
\end{equation} 

This differential equation cannot be analytically solved, so we must use numerical methods to approximate its solution. 

\subsection{Dimensionless Variables}

For notational simplicity we will drop the subscript $\theta$ in $\phi_\theta$ (recall $\theta$ represents parameters that are held fixed in the solution) and solve Eq.~\eqref{eq:poisson-boltzmann}   with boundary conditions Eqs.~\eqref{eq:boundary-plasma-radius2} and \eqref{eq:boundary-condition-origin-repeat}.

The first step in the numerical solution is transforming into dimensionless variables $(x,\phi,f)$:

\begin{align}
    x &:= \frac{r}{\lambda_\mathrm{D}}, \\
    \psi (x; \Vb) &:= \frac{q[\Vb - \phi(x \lambda_\mathrm{D}; \Vb)]}{ k_\mathrm{B} T}, \\
    f(x; \Vb) &:= \frac{n(x \lambda_\mathrm{D}; \Vb) }{n_0} = \mathrm{erf} \sqrt{\psi(x; \Vb)}. \label{eq:cumulative-definition}
\end{align}
Here, $x$ represents the radius in units of Debye length $\lambda_\mathrm{D}$, $\psi(x; \Vb)$ the barrier height at $x$ in units of $k_\mathrm{B} T$, and $f(x; \Vb)$ the fraction of initial particles remaining in the plasma at normalized radius $x$. Our definition of $\psi$ differs from Beck's definition (cf., Eq.~(4.46) in \cite{beck1990thesis}), and is more convenient for numerical solutions. 

Equation Eq.~\eqref{eq:poisson-boltzmann} transforms into

\begin{equation} \label{eq:poisson-boltzmann-dimensionless}
    \frac{1}{x} \frac{d}{dx} \bigg( x \frac{d\psi(x; \Vb)}{dx} \bigg) = \mathrm{erf} \sqrt{\psi(x; \Vb)},
\end{equation}
with one boundary condition

\begin{align} \label{eq:boundary-azimuthal-symmetry-dimensionless}
    \frac{d\psi(0; \Vb)}{dx} = 0 
\end{align}
at the axial center $x = 0$, and a second determined by requiring
\begin{align} \label{eq:boundary-plasma-radius-dimensionless}
  \psi(x_\mathrm{p}; \Vb) + x_\mathrm{p} \ln \bigg( \frac{R_\mathrm{w}}{R_\mathrm{p}} \bigg) \frac{d\psi(x_\mathrm{p}, \Vb)}{dx}  = \frac{q\Vb}{k_\mathrm{B}T}
\end{align}
at the dimensionless plasma radius $x = x_\mathrm{p} = R_\mathrm{p} / \lambda_\mathrm{D}$.
 We provide code for an integration algorithm at \\  \texttt{https://github.com/adriannez/nonneutral-plasma-temperature/} .

After numerically solving for $\psi(x; \Vb)$ for $x \in [0, x_\mathrm{p}]$, we   evaluate $\Nesctheta(\Vb) $  using Eq.~\eqref{eq:cumulative-definition-new}:
\begin{align} \label{eq:N-esc-from-psi}
  \Nesctheta(\Vb) &= N_0 - N_\theta(\Vb) \\ 
  &= N_0 + \bigg( \frac{2\pi \ell_\mathrm{p} \epsilon_0 R_\mathrm{p}}{q} \bigg) \frac{d\phi(r)}{dr}  \bigg|_{r = R_\mathrm{p}} \nonumber  \\ 
  &= N_0 \bigg[ 1 -  \bigg( \frac{2}{x_\mathrm{p}} \bigg) \frac{d\psi(x_\mathrm{p}; \Vb)}{dx}\bigg]. 
\end{align}
Here we used Eq.~\eqref{eq:cumulative-definition-new} in the second line, and the definition of $\lambda_\mathrm{D}$ from Eq.~\eqref{eq:lambda-debye} in the third.

It can be shown from Eq.~\eqref{eq:cumulative-definition} that: 
\begin{align} \label{eq:N-cyl-relation}
    \Nesctheta(\Vb) &= 2\pi l_p \int_0^{R_\mathrm{p}} n_0 [1 - f(r/ \lambda_\mathrm{D}; \Vb)] r \, dr \nonumber \\ 
    &= 2 n_0 \pi  \lambda_\mathrm{D}^2 l_p \int_0^{x_\mathrm{p}} \mathrm{erfc}\sqrt{ \psi(x; \Vb)} \, x \, dx \nonumber \\ 
    &= 2 N_\mathrm{cyl} \int_0^{x_\mathrm{p}}  \mathrm{erfc}\sqrt{ \psi(x; \Vb)} \, x  \, dx ,
\end{align}
where $\mathrm{erfc}(\cdot) = 1 - \mathrm{erf}(\cdot)$ denotes the complementary error function. The number of particles in a Debye cylinder $N_\mathrm{cyl} \propto T \ell_\mathrm{p}$ determines the  scaling of the escape function.

The mixed boundary conditions at $x = 0$ and $x = x_\mathrm{p}$ are satisfied by employing a shooting method  in which the Neumann boundary condition Eq.~\eqref{eq:boundary-azimuthal-symmetry-dimensionless} at $x = 0$ is supplanted with a guessed value for $\psi_\sigma(0)$, where $\sigma$ parameterizes the guessed solution. The initial conditions for both $\psi_\sigma(0)$ and $\psi_\sigma'(0)$ at $x = 0$ allow for the numerical solution for $\psi_\sigma(x)$, from $x=0$ to $x = x_\mathrm{p}$. 

The values for $\psi_\sigma(x_\mathrm{p})$ and $\psi'_\sigma(x_\mathrm{p})$ are used to evaluate the ``shot function''

\begin{equation} \label{eq:shot-guess}
  S (\sigma; \Vb) = \psi_\sigma (x_\mathrm{p}) + x_\mathrm{p} \ln \bigg( \frac{R_\mathrm{w}}{R_\mathrm{p}} \bigg) \frac{d\psi_\sigma(x_\mathrm{p}) }{dx} - \frac{q\Vb}{k_\mathrm{B}T}. 
\end{equation}
For a particular value of $\Vb$, the value of $\sigma^*$ that satisfies $S(\sigma^*; \Vb) = 0$ corresponds to the desired $\psi_{\sigma^*}(x) =  \psi(x; \Vb)$ that satisfies the boundary condition Eq.~\eqref{eq:boundary-plasma-radius-dimensionless}.

\subsection{Naive Lagrangian Discretization} 

In this section, we sketch  the Lagrangian-based discretization used in the integration of Eq.~\eqref{eq:poisson-boltzmann-dimensionless}. The discretization of Eq.~\eqref{eq:poisson-boltzmann-dimensionless} means that we discretize the $x$-axis into a grid of step-size $\Delta x$, with grid-points $x_n = n \Delta x$, with $n = 0, 1, ..., \lfloor x_p / \Delta x \rfloor$ and approximate the differential equation Eq.~\eqref{eq:poisson-boltzmann-dimensionless} with difference equations that propagate $(\psi(x_n), \psi'(x_n))$ forward to $(\psi(x_{n+1}), \psi'(x_{n+1}))$. Then, to numerically integrate from   initial conditions $(\psi(x_i), \psi'(x_i))$ at $x = x_i$ to $x = x_\mathrm{p}$, we make a forward propagation step of size $\Delta x_i < \Delta x$ to the closest grid-point $x_{n_i}$, iterate forward propagation steps of size $\Delta x$ till $x = x_f = \Delta x \lfloor x_p / \Delta x \rfloor$, and  make a final forward propagation step of size $\Delta x_f = x_\mathrm{p} - x_f$ to obtain $(\psi(x_\mathrm{p}), \psi'(x_\mathrm{p}))$.

The second-order differential equation Eq.~\eqref{eq:poisson-boltzmann-dimensionless} has a corresponding Lagrangian: 
\begin{equation} \label{eq:lagrangean-simple}
    L (\psi, \psi' , x) = \frac{x}{2} \bigg[ \bigg(\frac{\psi'^2}{2}\bigg) - V(\psi) \bigg]
\end{equation}
where the potential $ V(\psi)$ satisfies $-dV/d\psi = \mathrm{erf} \sqrt{\psi}$. We can then define the conjugate momentum $\Pi$ of Lagrangian Eq.~\eqref{eq:lagrangean-simple} to be
\begin{equation} \label{eq:conj-momentum-simple}
  \Pi(\psi, \psi', x) = \frac{\partial L}{\partial \psi'} = \bigg(\frac{x}{2}\bigg) \psi'.
\end{equation}
We define the discretized $\psi$ and $\Pi$ values at integer steps of the $x$ grid by: 
\begin{align*}
    \psi_n = \psi(x_n) \quad\mathrm{and}\quad \Pi_n = \Pi(x_n)
\end{align*}
and the position and slope $\psi' = d\psi / dx$ at half-integer steps by 

\begin{align*}
    x_{n+1/2} &= \frac{x_{n+1} + x_{n}}{2},
\end{align*}
and
\begin{align*}
    \psi'_{n+1/2} &= \frac{\psi_{n+1} - \psi_{n}}{\Delta x} .
\end{align*}

Following Ref.~\cite{symplectic-discretization}~(Sec. VI.6.2), we  discretize the Lagrangian Eq.~\eqref{eq:lagrangean-simple} to obtain difference equations that propagate $(\psi_n, \Pi_n)$ forward to $(\psi_{n+1}, \Pi_{n+1})$:
\begin{align} 
\psi'_{n+1/2} &= \bigg(\frac{2}{x_{n+1/2}}\bigg)\Pi_n + \frac{\Delta x}{2} \bigg(\frac{x_n}{x_{n+1/2}} \, \mathrm{erf} \sqrt{\psi_n} \bigg), \\ 
\psi_{n+1} &= \psi_n + \Delta x \, \psi'_{n+1/2}
\end{align}
and
\begin{align}
\Pi_{n+1} &= \Pi_n + \frac{\Delta x}{2} \bigg(\frac{x_n}{2} \, \mathrm{erf} \sqrt{\psi_n}  + \frac{x_{n+1}}{2} \, \mathrm{erf} \sqrt{\psi_{n+1}} \bigg).
\end{align}

Advantages of employing a symplectic integration method, in this case with Lagrangian-based discretization, are that the resulting flows from the discretized difference equations have an error of $o(\Delta x^2)$, as well as much better qualitative stability due to the preservation of certain geometric invariants \cite{symplectic-discretization}. 

See documented code at \texttt{https://www.github.com/adriannez/nonneutral-plasma- \\ temperature/} for further details about, e.g., how we implement initial conditions.

\section{Monte Carlo Simulations}

The Monte Carlo simulation starts with with parameters $\boldtheta = (T, n_0, R_\mathrm{p}, \ell_\mathrm{p})$ and a total of $N_0 = \lfloor n_0 \pi R_\mathrm{p}^2 \ell_\mathrm{p} \rfloor$ (discretized) plasma particles (indexed by $\alpha = 1, 2, ..., N_0$. Each particle $\alpha$ is given a randomly generated position $\mathbf{r}_\alpha$ from which we  calculate its potential energy $U_\alpha$ (by including the field created by  other particles and the applied potential). It has axial kinetic energy $K_\alpha = mv_{\parallel, \alpha}^2 / 2$. 

The particles are assigned a random axial velocity $v_{\parallel, \alpha}$ chosen as follows. The temperature of the plasma  $T$ is determines the distribution of kinetic energies by the one-dimensional Maxwellian distribution: $f(K) dK \propto \mathrm{exp}(-K/k_\mathrm{B} T) \sqrt{K} dK$. For each of the particles $\alpha$, we randomly assign an axial kinetic energy value $K_\alpha$ from this distribution. 

From the standpoint of computational complexity, once the positions of the $N_0$ particles are generated, the calculation of the potentials for the $N_0$ particles has a runtime of $O(N_0^2)$. This becomes prohibitively computationally expensive, as the plasmas we simulate have  $N_0 \approx O(10^{6})$ particles.

For the sake of computational tractability, we approximate the potential due to each particle as if it were a cylindrical shell with radius $r_\alpha = \sqrt{x_\alpha^2 + y_\alpha^2}$. Then, the potential energy of each particle $\alpha$ is:

\begin{align}
    U_\alpha &= q \sum_{\beta \neq \alpha}  \frac{q}{2\pi l_p \epsilon_0} \ln \bigg(\frac{R_w}{\max (r_\alpha, r_\beta)}\bigg)  \\ 
    &= \frac{q^2 N_{r < r_\alpha}}{2\pi l_p \epsilon_0} \ln \bigg(\frac{R_w}{r_\alpha}\bigg) + \sum_{\beta \, \mathrm{s.t.} \, r_\beta > r_\alpha} \frac{q^2}{2\pi l_p \epsilon_0} \ln \bigg(\frac{R_w}{r_\beta}\bigg)
\end{align}
as each cylindrical shell $\beta$ with larger radius $r_\beta > r_\alpha$ contributes to a term proportional to $\ln (R_w/R_\beta)$, while each cylindrical shell $\gamma$ with smaller radius $r_\gamma > r_\alpha$ contributes to a term proportional to $\ln (R_w/R_\alpha)$. Here, $N_{r < r_\alpha}$ denotes the number of plasma particles whose radii are smaller than $r_\alpha$. 

The particles' potential energies are found by sorting their radii in increasing order $\{r_1, r_2, ..., r_{_{N_0}}\}$, and then calculating, starting with the particle with the largest radius $r_{_{N_0}}$:

\begin{equation}
    U_{_{N_0}} = \frac{q^2(N_0 - 1)}{2\pi l_p \epsilon_0} \ln \bigg(\frac{R_w}{r_{_{N_0}}}\bigg),
\end{equation}
and iterating towards lower radii:
\begin{equation}
    U_{k-1} = U_{_{k}} + \frac{q^2(k - 2)}{2\pi l_p \epsilon_0} \ln \bigg(\frac{r_{k}}{r_{k-1}}\bigg),
\end{equation}
where $N_{r < r_{k - 1}} = k - 2$. The computational runtime when sorting is $O(N_0 \log N_0)$. 
The total particle energy is  $E_\alpha = K_\alpha + U_\alpha$. 

At any given moment, with $N$ remaining plasma particles (starting with $N = N_0$), we find the particle $\alpha$ with the highest energy $E_\alpha$,  remove it from the plasma and record its corresponding escape potential $V_\alpha = E_\alpha / q$. (This corresponds to when the barrier voltage $\Vb$ is slowly lowered to a value that allows the most energetic particle $\alpha$ in the remaining plasma to escape, and then recording $\Vb = E_\alpha / q$.) For each remaining particle $\beta$, its potential energy $U_\beta$ is lowered by: 

\begin{equation} \label{eq:voltage-update-rule}
    U_\beta \leftarrow U_\beta - \frac{q^2}{2\pi l_p \epsilon_0} \ln \bigg(\frac{R_w}{\max (r_\alpha, r_\beta)}\bigg).
\end{equation}
Since the particles are  sorted by increasing radius, we only need to update $U_\mathrm{\beta}$  on average $N / 2$ times per escape. This step is repeated until the desired number of of particles escape  (corresponding to $\Ndata$), ultimately giving a computational runtime $O(N_{0} \Ndata)$.

The $\Ndata$ escape voltages $\{V_\alpha \}$  correspond to a single instance of a plasma with parameters $\boldtheta$.  One simulation has  computational runtime $O(N_0 \log N_0) + O(N_{0} \Ndata)$ from the plasma generation and plasma steps, respectively. This is repeated $S$ times to get an ensemble of $S$ sets of escape voltages. Each instance is then fed into the temperature diagnostic algorithm to obtain the Monte Carlo ensemble estimated parameters $\{\hat{\boldtheta}_s\}$.

\section{Priors and Degeneracy of Fit} 

We found that for certain parameter regimes the fitter is underconstrained, i.e., there are multiple parameters yielding nearly the same fit (all with the same temperature $\pm \ 2$\%). To resolve this, we implement experimentally-motivated priors on a subset of our four parameters $\boldtheta$. (Experimentally, the plasma length $\ell_\mathrm{p}$ is an easily manipulated parameter and total charge $Q_0$ may be readily measurable using a Faraday cup, yielding $N_0 = Q_0 / q$.) 

We reduce our parameter-space to a two-parameter fit using the equality constraints

\begin{equation}
     \hat{\ell}_\mathrm{p} = \ell_\mathrm{p},
\end{equation}
and
\begin{equation}
    \hat{n}_0 \hat{R}_\mathrm{p}^2 = \frac{N_0}{\pi \ell_\mathrm{p}},
\end{equation}
where the plasma length $\ell_\mathrm{p}$ and total particle count $N_0 = n_0 \pi R_\mathrm{p}^2 \ell_\mathrm{p}$ are assumed to be known exactly or within a small range of values. Given these two constraints, we use the reduced parameters $\boldtheta_\mathrm{reduced} = (\hat{T}, \hat{\phi}_\mathrm{center})$, with $\phi_\mathrm{center}$ corresponding to $\phi_0(r = 0)$: 

\begin{align}
    \hat{\phi}_\mathrm{center} = \frac{q N_0}{4 \pi \epsilon_0 \ell_\mathrm{p}} \bigg( 1 + 2 \ln \bigg(\frac{R_\mathrm{w}}{\hat{R}_\mathrm{p}} \bigg)\bigg). 
\end{align}
We find that this parameter reduction resolves all underconstrained cases.

\end{document}